\documentclass[11pt,preprintnumbers,aps,amssymb,nofootinbib,amsmath,superscriptaddress]
{revtex4}
\usepackage{epsfig,epsf}
\usepackage{xfrac} 
\usepackage{bm} 
\newcommand{\beq}{\begin{equation}}
\newcommand{\beql}[1]{\begin{equation}\label{#1}}
\newcommand{\eeq}{\end{equation}}
\newcommand{\eq}[1]{(\ref{#1})}

\renewcommand{\sec}[1]{Sec.~\ref{#1}}
\newcounter{topiccounter}
\renewcommand{\b}[1]{{\bm #1}} 

\newcommand{\as}{\alpha_s}

\newcommand{\e}{\varepsilon}

\begin{document}

\title{On  viscous flow and azimuthal anisotropy of quark-gluon plasma in strong magnetic field}

\author{Kirill Tuchin}

\affiliation{Department of Physics and Astronomy, Iowa State University, Ames, IA 50011}

\date{\today}


\begin{abstract}

We calculate the viscous pressure tensor  of the quark-gluon plasma in strong magnetic field. It is azimuthally anisotropic and is characterized by five shear viscosity coefficients, four of which vanish when the field strength $eB$ is much larger than the plasma temperature squared. We argue, that the azimuthally anisotropic viscous pressure tensor generates the transverse flow with  asymmetry as large as 1/3, even not taking into account the collision geometry. We conclude, that the magnitude of the shear viscosity extracted from the experimental data ignoring the magnetic field  must be  underestimated. 

\end{abstract}

\maketitle

\section{introduction}\label{sec:intro}

Strong magnetic field produced in relativistic  heavy-ion collisions \cite{Kharzeev:2007jp,Skokov:2009qp} has a strong impact on phenomenology of the quark-gluon plasma (QGP). It induces energy loss by fast quarks and charged leptons via the synchrotron radiation  \cite{Tuchin:2010vs} and  polarization of the fermion spectra  \cite{Tuchin:2010vs}. It contributes to the   enhancement of the dilepton production \cite{Tuchin:2010gx} and  azimuthal anisotropy of the quark-gluon plasma (QGP) \cite{Mohapatra:2011ku}.  It causes  dissociation of the bound states, particularly charmonia,  via ionization  \cite{Marasinghe:2011bt,Tuchin:2011cg}.
Additionally, the magnetic field drives the Chiral Magnetic Effect (CME) \cite{Kharzeev:2004ey,Kharzeev:2007jp,Kharzeev:2007tn,Fukushima:2008xe,Kharzeev:2009fn}, which is the generation of an electric field parallel to the magnetic one via the axial anomaly in the hot nuclear matter. 

It has been argued recently in \cite{Mohapatra:2011ku} that the magnetic field of strength $eB\simeq m_\pi^2$ \cite{Kharzeev:2007jp,Skokov:2009qp} is able to induce the azimuthal anisotropy of the order of $30\%$ on produced particles. This conclusion was reached by utilizing the solution of the magneto-hydrodynamic equations in weak magnetic field. In this paper we discuss the magneto-hydrodynamics of the QGP in the limit of strong magnetic field. Our goal is to calculate the effect of the magnetic field on viscosity of the plasma. It is well-known that the viscous pressure tensor of  magnetoactive plasma is characterized by seven viscosity coefficients,  among which five are shear viscosities and two are bulk ones. Generally, calculation of the viscosities requires knowledge of the strong interaction dynamics of the QGP components. However, in strong enough magnetic field  these interactions can be considered as a perturbation and viscosities can be analytically calculated using the kinetic equation.  Application of this approach to the non-relativistic electro-magnetic plasma is discussed in \cite{LLX}. A general relativistic approach was developed in \cite{Erkelens:1977}. We apply it in \sec{sec:kinetic} to derive the viscosity coefficients of QGP, which are given by \eq{visc3} and \eq{visc1}. As in the non-relativistic case, we found that four viscosities vanish as the magnetic field strength increases. 

A characteristic feature of the viscous pressure tensor in magnetic field is its azimuthal anisotropy. This anisotropy is the result of  suppression of the momentum transfer in QGP in the direction perpendicular to the magnetic field. Its macroscopic manifestation is decrease of the viscous pressure tensor components in the plane perpendicular to the magnetic field, which coincides with the ``reaction plane" in the heavy-ion phenomenology. Since Lorentz force vanishes in the direction parallel to the field,  viscosity along that direction is not affected at all. In fact, the viscous pressure tensor component in the reaction plane is twice as small as the one in the field direction. As the result,  transverse flow of QGP develops azimuthal anisotropy in presence of the magnetic field. Clearly, this anisotropy is completely different from the one generated by the anisotropic pressure gradients and exists even if the later are absent. 

In \sec{sec:flow} we discuss QGP transverse flow in strong magnetic field using the Navie-Stokes equations. At later times after the heavy-ion collision, flow velocity is proportional to  $\eta^{-1/2}$, see  \eq{sol1} and \eq{sol2}. If the system is such that in absence of the magnetic field it were azimuthally symmetric, then the magnetic field induces azimuthal asymmetry of 1/3, see \eq{anisotropy}. This is surprisingly close to the weak field limit recently reported in \cite{Mohapatra:2011ku}. The effect of the magnetic field on flow is strong and must be taken into account in phenomenological applications. 
Neglect of the contribution by the magnetic field  leads to underestimation of the phenomenological value of viscosity extracted from the data \cite{Song:2007fn,Romatschke:2007mq,Dusling:2007gi}.  In other words, more viscous QGP in magnetic field produces the same azimuthal anisotropy as a less viscous QGP in vacuum.

\section{Viscous pressure in strong magnetic field}\label{sec:kinetic}

\subsection{Kinetic equation}
Kinetic equation for the distribution function $f$ of a quark flavor of charge $ze$ is 
\beq\label{kinetic0}
p^\mu \partial_\mu f= zeB^{\mu\nu}\frac{\partial f}{\partial u^\mu}u_\nu +\mathcal{C}[ f,\dots]
\eeq
where $\mathcal{C}$ is the collision integral and $B^{\mu\nu}$ is the electro-magnetic tensor, which contains only magnetic field components in the laboratory frame. Ellipsis in the argument of $\mathcal{C}$ indicates  the distribution  functions of other quark flavors and gluons (we will omit them below). 
The equilibrium  distribution:
\beq\label{f0}
f_0= \frac{\rho}{4\pi m^3 T K_2(\beta m)}e^{ - \beta\, p\cdot U(x)}
\eeq
where $U(x)$ is the macroscopic velocity of fluid, $p^\mu= m u^\mu$ is particle momentum, $\beta = 1/T$ and $\rho$ is the mass density. 
Since $\frac{\partial f_0}{\partial u^\mu}\propto u_\mu$, the first term on the r.h.s.\ of \eq{kinetic0} vanishes in equilibrium as well as the collision integral. Therefore, we can write the kinetic equation as an equation for $\delta\! f$
\beq\label{kinetic}
p^\mu \partial_\mu f_0= zeB^{\mu\nu}\frac{\partial (\delta\! f)}{\partial u^\mu}u_\nu +\mathcal{C}[\delta\! f]
\eeq
where $\delta\! f$ is a deviation from equilibrium. Differentiating  \eq{f0} we find 
\beq\label{derf0}
\partial_\mu f_0= -f_0\frac{1}{T}\,p^\lambda\partial_\mu U_\lambda(x)
\eeq
Since $U^\lambda= (\gamma_V, \gamma_V \b V)$ and $p^\lambda= (\e, \b p)=(\gamma_v m,\gamma_v m\b v)$ it follows 
\beq
p\cdot U= \frac{m}{\sqrt{1-v^2}\sqrt{1-V^2}}(1-\b v\cdot \b V)
\eeq
Thus, in the comoving frame
\beql{z1}
\partial_\mu f_0|_{\b V=0}= f_0 \,\frac{1}{T}\, p_\nu \partial_\mu V^\nu
\eeq
Substituting \eq{z1} in \eq{kinetic} yields
\beq\label{kinetic2}
-\frac{f_0}{T}p^\mu p^\nu V_{\mu\nu}= zeB^{\mu\nu}\frac{\partial (\delta\! f)}{\partial u^\mu}u_\nu +\mathcal{C}[\delta\! f]
\eeq
where we defined
\beq \label{Vdef}
V_{\mu\nu}= \frac{1}{2}(\partial_\mu V_\nu+\partial_\nu V_\mu)
\eeq
and used $u^\mu u^\nu \partial_{\mu}V_\nu = u^\mu u^\nu V_{\mu\nu}$.

 Since the time-derivative of $f_0$ is irrelevant for the calculation of the viscosity we will drop it from the kinetic equation. All indexes thus become the usual three-vector ones. To avoid confusion we will label them by Greek letters from the beginning of the alphabet.  Introducing $b_{\alpha\beta}= B^{-1}\varepsilon_{\alpha\beta\gamma}B_\gamma$ we cast \eq{kinetic2} in the form
\beq\label{ke}
\frac{1}{T}p^\alpha u^\beta V_{\alpha\beta}f_0=  - zeB b_{\alpha\beta} v^\beta \frac{\partial (\delta\! f)}{\partial v_\alpha} \frac{1}{\e}-\mathcal{C}[\delta\! f] \,.
\eeq

The viscous pressure  generated by a deviation from equilibrium is given by the tensor
\beq\label{stress-def}
-\Pi_{\alpha\beta}= \int p_\alpha p_\beta\, \delta\! f \,\frac{d^3p}{\e}
\eeq
Effectively it can be parameterized in terms of the viscosity coefficients as follows (we neglect bulk viscosities)
\beq\label{visc.tens1}
\Pi_{\alpha\beta} = \sum_{n=0}^4 \eta_n\, V^{(n)}_{\alpha\beta}
\eeq
where the linearly independent tensors $V^{(n)}_{\alpha\beta}$ are given by
\begin{subequations}\label{vns}
\begin{eqnarray}\label{visc.tens}
V^{(0)}_{\alpha\beta} &=& \left(3b_\alpha b_\beta-\delta_{\alpha\beta}\right) \left(b_\gamma b_\delta V_{\gamma\delta}-\frac{1}{3}\nabla\cdot \b V\right) \\
V^{(1)}_{\alpha\beta}&=& 2V_{\alpha\beta}+
\delta_{\alpha\beta}V_{\gamma\delta}b_\gamma b_\delta  - 2V_{\alpha\gamma}b_\gamma b_\beta- 2V_{\beta\gamma}b_\gamma b_\alpha + (b_\alpha b_\beta-\delta_{\alpha\beta}) \nabla\cdot \b V +b_\alpha b_\beta V_{\gamma\delta} b_\gamma b_\delta\\
V^{(2)}_{\alpha\beta}&=& 2(V_{\alpha\gamma} b_{\beta\gamma} +V_{\beta\gamma} b_{\alpha\gamma}- V_{\gamma\delta} b_{\alpha\gamma}b_\beta b_\delta )\\
V^{(3)}_{\alpha\beta}&=& V_{\alpha\gamma}b_{\beta \gamma} + V_{\beta\gamma} b_{\alpha\gamma} - V_{\gamma\delta} b_{\alpha\delta} b_{\alpha\gamma} b_\beta b_\delta - V_{\gamma\delta} b_{\beta \gamma} b_\alpha b_\delta\\
V^{(4)}_{\alpha\beta}&=& 2( V_{\gamma\delta} b_{\alpha\delta} b_{\alpha\gamma} b_\beta b_\delta+
V_{\gamma\delta} b_{\beta \gamma} b_\alpha b_\delta)\,.
\end{eqnarray}
\end{subequations}
For calculation of shear viscosities $\eta_n$, $n=1,\ldots,4$ we can set $\nabla\cdot \b V=0$ and $V_{\alpha\beta}b_\alpha b_\beta=0$.

Let us expand $\delta\! f$  to the second order in velocities in terms of the tensors $V_{\alpha\beta}^{(n)}$ as follows
\beq\label{delf}
\delta\! f = \sum_{n=0}^4 g_n V^{(n)}_{\alpha\beta} v^\alpha v^\beta
\eeq
Then, substituting \eq{delf} into \eq{visc.tens1} and requiring consistency of \eq{stress-def} and \eq{visc.tens1} yields 
\beq\label{visc}
\eta_n = -\frac{2}{15}\int \e v^4 g_n d^3 p
\eeq 
This gives the viscosities in the magnetic field in terms of deviation of the distribution function from equilibrium. Transition to the  non-relativistic limit in \eq{visc} is achieved by the replacement $\e\to m$ \cite{LLX}.


\subsection{Viscosity of collisionless plasma}\label{sec:B}

In strong magnetic field we can determine $g_n$ by the method of consecutive approximations.
Writing $\delta\! f = \delta\! f^{(1)}+\delta\! f^{(2)}$ and substituting into \eq{ke} we find
\beq\label{ke-delta}
\frac{1}{T}p^\alpha v^\beta V_{\alpha\beta}f_0=  - zeB b_{\alpha\beta} v^\beta \frac{\partial (\delta\! f^{(1)}+\delta\! f^{(2)})}{\partial v_\alpha} \frac{1}{\e}+\mathcal{C}[\delta\! f^{(1)}] \,.
\eeq
Here we assumed that  the deviation from equilibrium due to the strong magnetic field is much larger than due to particle  collisions. The explicit form of $\mathcal{C}$ is determined by the strong interaction dynamics but drops off the equation in the leading oder. 
The first correction to the equilibrium distribution obeys the equation
\beq\label{deltaf1}
\frac{1}{T}p_\alpha v_\beta V_{\alpha\beta}f_0=  - zeB b_{\alpha\beta} v_\beta \frac{\partial \delta\! f^{(1)}}{\partial v_\alpha} \frac{1}{\e} \,.
\eeq
Using \eq{delf} we get
\beql{w1}
b_{\alpha\beta} v_\beta\frac{\partial \delta\! f^{(1)}}{\partial v_\alpha} = 2b_{\alpha\beta} v_\beta\sum_{n=0}^4 g_n\, V^{(n)}_{\alpha\gamma}v_\gamma
\eeq
Substituting \eq{w1} into \eq{deltaf1} and using \eq{vns} yields:
\begin{eqnarray}\label{cond1}
\frac{\e}{T zeB}\,p_\alpha v_\beta V_{\alpha\beta}f_0&=& - 2b_{\beta\nu} v_\alpha v_\nu [ g_1(2V_{\alpha\beta}-2 V_{\beta\gamma}b_\gamma b_\alpha)+2g_2 V_{\beta\gamma} b_\gamma b_\alpha\nonumber\\
&& +g_3 (V_{\alpha\gamma} b_{\beta\gamma}+V_{\beta\gamma}b_{\alpha\gamma}-V_{\gamma\delta}b_\alpha b_\delta)+2g_4 V_{\gamma\delta} b_{\beta\gamma} b_\alpha b_\delta)]
\end{eqnarray}
where we used the following identities $b_{\alpha\beta}b_\alpha =b_{\alpha\beta}b_\beta=b_{\alpha\beta}v_{\alpha}v_\beta=0$. Clearly, \eq{cond1} is satisfied only if $g_1=g_2=0$. Concerning the other two coefficients, we use the identities
\begin{subequations}
\begin{eqnarray}
b_{\alpha\beta}b_{\beta\gamma}&=& b_\gamma b_\alpha - \delta_{\alpha\gamma}b^2\,,\\ 
  \varepsilon_{\alpha\beta\gamma}\varepsilon_{\delta\epsilon\zeta}    &  =& \delta_{\alpha\delta}\left( \delta_{\beta\epsilon}\delta_{\gamma\zeta} - \delta_{\beta\zeta}\delta_{\gamma\epsilon}\right) - \delta_{\alpha\epsilon}\left( \delta_{\beta \delta}\delta_{\gamma\zeta} - \delta_{\beta\zeta}\delta_{\gamma\delta} \right) + \delta_{\alpha\zeta} \left( \delta_{\beta\delta}\delta_{\gamma\epsilon} - \delta_{\beta\zeta}\delta_{\gamma\delta} \right) \,
\end{eqnarray}
\end{subequations}
that we substitute into \eq{cond1} to derive
\beql{prom1}
-\frac{\e}{2T zeB}\,p^\alpha v^\beta V_{\alpha\beta}f_0=  g_3 [2V_{\alpha\beta} b_\alpha b_{\beta}-4V_{\alpha\beta}v_\alpha b_\beta (\b b\cdot\b  v)]+2g_4 V_{\alpha\beta}v_\alpha b_\beta (\b b\cdot\b  v)\,.
\eeq
Since $p_\alpha = \e v_\alpha$ we obtain
\beql{g3g4}
g_3 = \frac{g_4}{2}= -\frac{\e^2 f_0}{4T zeB}
\eeq

Using \eq{f0}, \eq{g3g4} in \eq{visc} in the comoving frame (of course $\eta_n$'s do not depend on the frame choice) and integrating using 3.547.9 of \cite{GR} we get 
\beql{visc3}
\eta_3= \frac{K_3(\beta m)}{K_2(\beta m)}\frac{\rho T}{2  zeB}
\eeq
The non-relativistic limit corresponds to $m\gg T$ in which case we get
\beql{eta3NR}
\eta_3^\mathrm{NR}=\frac{\rho T}{2zeB}\,.
\eeq
In the opposite ultra-relativistic case $m\ll T$ (high-temperature plasma)
\beql{eta3UR}
\eta_3^\mathrm{UR}=\frac{2 n T^2   }{zeB}\,.
\eeq
where $ n= \rho/m$ is  the number density.

\subsection{Contribution of collisions}

In the relaxation-time approximation we can write the collision integral as 
\beql{rel.time}
\mathcal{C}[\delta\! f]= -\nu\, \delta\! f
\eeq
where $\nu$ is an effective collision rate. Strong field limit means that
\beql{con}
\omega_B\gg \nu
\eeq
where $\omega_B = zeB/\e$ is the synchrotron frequency. Whether $\nu$ itself is function of the field depends on the relation between the Larmor radius $r_B=v_T/\omega_B$, where $v_T$ is the particle velocity in the plane orthogonal to $\b B$ and the  Debye radius $r_D$. If 
\beql{w}
r_B\gg r_D
\eeq
then the effect of the field on the collision rate  $\nu$ can be neglected  \cite{LLX}. Assuming that \eq{w} is satisfied the collision rate reads
\beql{nu}
\nu = nv\sigma_t
\eeq
where $\sigma_t$ is the transport cross section,  which is a function of the saturation momentum $Q_s$ \cite{Gribov:1984tu,Blaizot:1987nc}. We estimate $\sigma_t\sim \as^2 /Q_s^2$, with $Q_s\sim 1$~GeV and $n= P/T$ with pressure $\as^2P\sim 1$~GeV/fm$^3$ we get $\nu\sim 40$~MeV. Inequality \eq{con} is well satisfied since $eB\simeq m_\pi^2$  \cite{Kharzeev:2007jp,Skokov:2009qp} and $m$ is  in the range between the current and the constituent quark masses. On the other hand, applicability of the condition \eq{w} is marginal and is very sensitive to the interaction details. In this section we assume that \eq{w} holds in order to obtain the analytic solution. Additionally, the general condition for the applicability of the hydrodynamic approach $\ell = 1/\nu \ll L$, where $\ell$ is the mean free path and $L$ is the plasma size is assumed to hold. Altogether we have $r_D\ll r_B \ll \ell \ll L$.
 
Equation for the second correction to the equilibrium distribution $\delta\! f^{(2)}$ follows from \eq{ke-delta} after substitution \eq{rel.time}
\beql{second-cor}
\frac{zeB}{\e} b_{\alpha\beta} v_\beta \frac{\partial \delta\! f^{(2)}}{\partial v_\alpha} = - \nu \delta\! f^{(1)}
\eeq
Now, plugging 
\begin{subequations}\label{df1}
\begin{eqnarray}
\delta\! f^{(1)}&=& [g_3 V_{\alpha\beta}^{(3)}+g_4 V_{\alpha\beta}^{(4)}] v_\alpha v_\beta\,,\\
\delta\! f^{(2)}&=& [g_1 V_{\alpha\beta}^{(1)}+g_2 V_{\alpha\beta}^{(2)}] v_\alpha v_\beta
\end{eqnarray}
\end{subequations}
into \eq{second-cor} yields
\begin{eqnarray}
\frac{2zeB}{\e}\left\{ g_1[ 2V_{\beta\alpha}b_{\alpha\gamma}v_\beta v_\gamma - 
2V_{\beta\alpha}b_{\alpha\gamma}v_\beta v_\gamma (\b v\cdot \b b)]+2g_2 V_{\beta\alpha}b_{\alpha\gamma}v_\beta v_\gamma (\b v\cdot \b b)\right\}&&\nonumber\\
=-\nu g_3\left \{ -2 V_{\beta\alpha}b_{\alpha\gamma}v_\beta v_\gamma - 6 V_{\beta\alpha}b_{\alpha\gamma}v_\beta v_\gamma (\b v\cdot \b b)\right\}
\end{eqnarray}
where we used $g_4=2g_3$. It follows that 
\beql{g1g2}
g_1= \frac{g_2}{4}= \frac{\nu\gamma_v g_3}{2\omega_B} 
\eeq
With the help of \eq{nu},\eq{f0},\eq{visc} we obtain
\beql{visc1}
\eta_1=\frac{\eta_2}{4}= \frac{8}{5\sqrt{2\pi}}\frac{\rho^2\sigma_t \,T^{3/2}}{(zeB)^2 m^{1/2}}\frac{K_{7/2}(\beta m)}{ K_2(\beta m)}
\eeq

\section{Transverse flow}\label{sec:flow}

To illustrate the effect of the magnetic field on the viscous flow of the electrically charged component of the quark-gluon plasma we will assume that the flow is non-relativistic and use the Navie-Stokes equations that read
\beql{n-s}
\rho \left( \frac{\partial V_\alpha}{\partial t}+V_\beta\frac{\partial V_\alpha}{\partial x_\beta}\right)= -\frac{\partial P}{\partial x_\alpha}+\frac{\partial \Pi_{\alpha\beta}}{\partial x_\beta}
\eeq
where $\Pi_{\alpha\beta}$ is the viscous pressure tensor, $\rho = mn$ is mass-density and $P$ is pressure. We will additionally 
assume that the flow is non-turbulent and that the plasma is non-compressible. The former assumption amounts to dropping the non-linear in velocity terms, while the later implies vanishing divergence of velocity
\beql{non-comp}
\nabla\cdot \b V=0
\eeq 
Because of the approximate boost invariance of the heavy-ion collisions, we can restrict our attention to the two dimensional flow in the $xz$ plane corresponding to the central rapidity region. 

The viscous pressure tensor in vanishing magnetic field is isotropic in the $xz$-plane and is given by 
\beql{P0}    
\Pi_{\alpha\beta}^0= \eta \left( \frac{\partial V_\alpha}{\partial x_\beta}+\frac{\partial V_\beta}{\partial x_\alpha}\right) = 2\eta \left(
\begin{array}{cc}
V_{xx} & V_{xz}\\
V_{zx} & V_{zz}
\end{array}
\right)
\eeq
where the superscript $0$ indicates absence of the magnetic field. In the opposite case of very strong magnetic field the viscous pressure tensor has a different form \eq{visc.tens1}. Neglecting all $\eta_n$ with $n\ge 1$ we can write
\beql{PB}
\Pi_{\alpha\beta}^\infty = \eta_0\left(
\begin{array}{cc}
-V_{zz} & 0\\
0 & 2V_{zz}
\end{array}
\right) = 
2\eta_0\left(
\begin{array}{cc}
\frac{1}{2}V_{xx} & 0\\
0 & V_{zz}
\end{array}
\right)
\eeq
where we also used \eq{non-comp}. 
Notice that $\Pi_{xx}^\infty=\frac{1}{2}\Pi_{zz}^\infty=\frac{1}{2}\Pi_{xx}^0$ indicating that the plasma flows in the direction perpendicular to the magnetic field with twice as small viscosity as in the direction of the field. The later is not affected by the field at all, because the Lorentz force vanishes in the field direction.
Substituting \eq{PB} into \eq{n-s} we derive the following two equations characterizing the plasma velocity in strong magnetic field  
\begin{align}\label{eq1}
\rho \frac{\partial V_x}{\partial t}= -\frac{\partial P}{\partial x}+\eta_0\frac{\partial ^2 V_x}{\partial x^2}\,,\qquad 
\rho \frac{\partial V_z}{\partial t}= -\frac{\partial P}{\partial z}+2\eta_0\frac{\partial ^2 V_z}{\partial z^2}
\end{align}
Additionally we need to set two initial conditions 
\begin{align}\label{ic1}
V_x\big|_{t=0}= \varphi_1(x,z)\,, \qquad V_z\big|_{t=0}= \varphi_2(x,z)
\end{align}
The solution to the the problem \eq{eq1},\eq{ic1} is
\begin{subequations}
\begin{align}\label{sol1}
&V_x(x,z,t)= \int_{-\infty}^\infty dx' \varphi_1(x',z)G_{\frac{1}{2}}(x-x',t)-\frac{1}{\rho}\int_0^tdt'\int _{-\infty}^\infty dx' G_{\frac{1}{2}}(x-x',t-t')\frac{\partial P(x',z,t') }{\partial x'}\\
&V_z(x,z,t)= \int_{-\infty}^\infty dz' \varphi_2(x,z')G_1(z-z',t)-\frac{1}{\rho}\int_0^tdt'\int _{-\infty}^\infty dz' G_1(z-z',t-t')\frac{\partial P(x,z',t') }{\partial z'}\label{sol2}
\end{align}
\end{subequations}
Here the Green's function is given by 
\beql{Green}
G_k(z,t)= \frac{1}{\sqrt{4\pi a^2k t}}e^{-\frac{z^2}{4a^2kt}}
\eeq
and the diffusion coefficient by
\beql{dif}
a^2= \frac{2\eta_0}{\rho}
\eeq

Suppose that the pressure is isotropic, i.e.\ it depends on the coordinates $x$,$z$ only via the radial coordinate $r= \sqrt{x^2+z^2}$; accordingly we pass from the integration variables $x'$  and $z'$ to $r$  in \eq{sol1} and \eq{sol2} correspondingly.  At later times we can expand the Green's function \eq{Green} in inverse powers of $t$. The first terms in the r.h.s.\ of \eq{sol1} and \eq{sol2} are subleasing  and we obtain
\begin{subequations}
\begin{align}\label{q1}
V_x(x,z,t)&\approx  -\frac{1}{\rho}\int_0^tds \int _{-\infty}^\infty dr\, \frac{1}{\sqrt{2\pi a^2 s}}\frac{\partial P(r,t-s) }{\partial r}\nonumber\\
&=-\frac{1}{\rho}\int_0^tds  \frac{1}{\sqrt{2\pi a^2 s}}\left[ P(R(s),t-s)-P(0,t-s)\right] 
\end{align}
and by the same token 
\begin{align}\label{q2}
&V_z(x,z,t)\approx -\frac{1}{\rho}\int_0^tds \, \frac{1}{\sqrt{4\pi a^2 s}}\left[ P(R(s),t-s)-P(0,t-s)\right]
\end{align}
\end{subequations}
where $R(t)$ is the boundary beyond which the density of the plasma is below the critical value. 
We observe that $V_x/V_z= \sqrt{2}$.
Consequently, the azimuthal anisotropy of the hydrodynamic flow is
\beql{anisotropy}
\frac{V_x^2-V_z^2}{V_x^2+V_z^2}= \frac{1-\frac{1}{2}}{1+\frac{1}{2}}=\frac{1}{3}
\eeq
Since we assumed that the  initial conditions and  the pressure are isotropic, the azimuthal asymmetry \eq{anisotropy} is generated exclusively by the magnetic field. 

\section{summary}

The structure of the viscous stress tensor in a very strong magnetic field \eq{PB} is general, model independent. However the precise amount of the azimuthal anisotropy that it generates is of course model dependent. We however  draw the reader's attention to the fact that analysis of \cite{Mohapatra:2011ku} using quite different arguments arrived at a very similar estimate. Although a more quantitive numerical calculation is certainly required before a final conclusion can be made, it looks very plausible that the QGP viscosity is significantly higher than the presently accepted value extracted without taking into account the magnetic field effect \cite{Song:2007fn,Romatschke:2007mq,Dusling:2007gi} and is perhaps closer to the value calculated using the  perturbative theory \cite{Arnold:2003zc,Baym:1990uj}.

\acknowledgments
This work  was supported in part by the U.S. Department of Energy under Grant No.\ DE-FG02-87ER40371.



\end{document}